# On Determining the Fair Bandwidth Share for ABR Connections in ATM Networks

Sonia Fahmy, Raj Jain, Shivkumar Kalyanaraman, Rohit Goyal and Bobby Vandalore
The Ohio State University
E-mail: {*fahmy,jain*}@cis.ohio-state.edu

**Abstract:** The ABR service is designed to fairly allocate the bandwidth unused by higher priority services. The network indicates to the ABR sources the rates at which they should transmit to minimize their cell loss. Switches must constantly measure the demand and available capacity, and divide the capacity fairly among the contending connections. In order to compute the fair and efficient allocation for each connection, a switch needs to determine the effective number of active connections. In this paper, we propose a method for determining the number of active connections and the fair bandwidth share for each. We prove the efficiency and fairness of the proposed method analytically, and simulate it for a number of configurations.

## 1 Introduction

ATM networks offer five service categories: constant bit rate (CBR), real-time variable bit rate (rt-VBR), non-real time variable bit rate (nrt-VBR), available bit rate (ABR), and unspecified bit rate (UBR). The ABR and UBR service categories are specifically designed for data traffic. The ABR service provides better service for data traffic than UBR by frequently indicating to the sources the rate at which they should be transmitting. For this reason, an ATM switch must compute the fair bandwidth share for each of the active ABR connections.

Determining the fair bandwidth share for the active ABR connections is an extremely complex problem. This is because fairness is commonly measured by the max-min fairness criteria (defined in the next section). Intuitively, fairness means that if a connection is bottlenecked elsewhere, it should be allocated the maximum it can use, and the left over capacity should be fairly divided among the connections that can use it. The switch should indicate this fair bandwidth share to the sources, while also accounting for the load and queuing delays at the switch.

This paper proposes a method to determine the fair bandwidth share for the active ABR connections, and analyzes the performance of this method using both simple mathematical proofs and simulations. The remainder of the paper is organized as follows. In the next section, we review the ABR flow control mechanisms in ATM networks. Then, we describe the original ERICA switch algorithm which is modified in this study. Sections 4 and 5 point out some problems with the original ERICA algorithm, and describe how ERICA has solved these problems. We then describe our proposed method (which also overcomes those problems), and give a proof of its correctness, and a number of examples of its operation. Finally, we analyze the performance of the proposed method.

## 2 ABR Flow Control

As previously mentioned, the ABR service frequently indicates to the sources the rate at which they should be transmitting. The feedback from the switches to the sources is indicated in Resource Management (RM) cells which are generated periodically by the sources and turned around by the destinations. Figure 1 illustrates this operation.

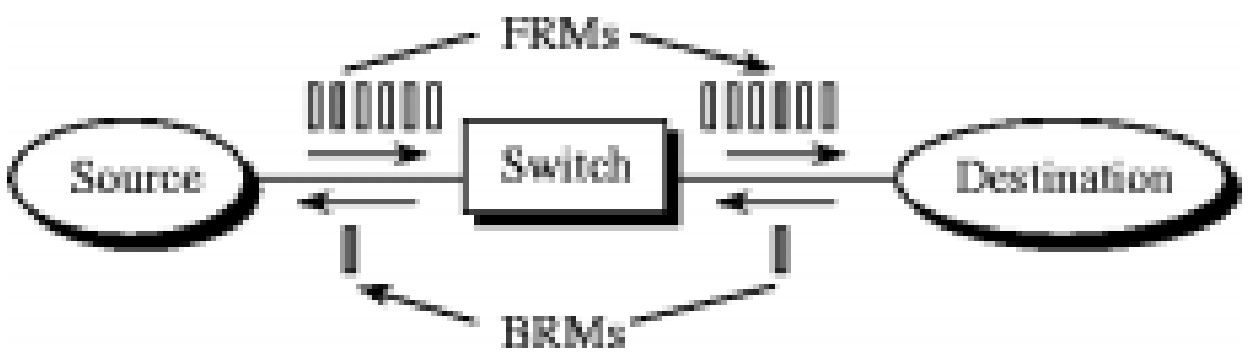

Figure 1: Resource management cells in an ATM network

The RM cells contain the source current cell rate (CCR), in addition to several fields that can be used by the switches to provide feedback to the sources. Among these fields is the explicit rate (ER) field, which indicates the rate that the network can support for this connection at that particular instant. At the source, the ER field is initialized to a rate no greater than the PCR (peak cell rate). Each switch on the path from the source to the destination reduces the ER field to the maximum rate it can support [5].[1]

The RM cells flowing from the source to the destination are called forward RM cells (FRMs) while those returning from the destination to the source are called backward RM cells (BRMs). When a source receives a BRM cell, it computes its allowed cell rate (ACR) using its current ACR value, and the ER field of the RM cell [7].

### 2.1 Fairness

The optimal operation of a distributed shared resource is usually given by a criterion called the *max-min allocation* [4].

[1] All our papers and ATM Forum contributions are available through http://www.cis.ohio-state.edu/~jain/

This fairness definition is the most commonly accepted one, though other definitions are also possible.

The max-min allocation is defined as follows. Given a configuration with $n$ contending sources, suppose the $i^{th}$ source is allocated a bandwidth $x_i$. The allocation vector $\{x_1, x_2, \ldots, x_n\}$ is feasible if all link load levels are less than or equal to 100%. Given an allocation vector, the source that is getting the least allocation is, in some sense, the "unhappiest source". We need to find the feasible vectors that give the maximum allocation to this unhappiest source. Now we remove this "unhappiest source" and reduce the problem to that of the remaining $n-1$ sources operating on a network with reduced link capacities. Again, we find the unhappiest source among these $n-1$ sources, give that source the maximum allocation and reduce the problem by one source. We repeat this process until all sources have been allocated the maximum that they can get.

# 3 Original ERICA Algorithm

Several switch algorithms have been developed to compute the feedback to be indicated to ABR sources in RM cells [1, 9, 10, 11, 8]. The ERICA algorithm [6, 8] is one of these algorithms. The main advantages of ERICA are its low complexity, fast transient response, high efficiency, and small queuing delay.

The ERICA algorithm aims at computing a fair and efficient allocation of the available bandwidth to all contending sources. In this section, we present the basic features of the original algorithm and explain their operation. The next sections describe some issues and additions to the algorithm, and a new method to determine the number of active connections. For a more complete description of the algorithm and its performance, refer to [8].

The ERICA switch periodically monitors the load on each link and determines a load factor, $z$, the available capacity, and the number of currently active virtual connections (VCs). The load factor is calculated as follows:

$$z \leftarrow \frac{\text{ABR Input Rate}}{\text{ABR Capacity}}$$

where:
ABR Capacity←Target Utilization $\times$ Link Bandwidth $-$ VBR Usage $-$ CBR Usage.

The input rate and output link ABR capacity are measured over an interval called the switch measurement interval. The above steps are executed at the end of the switch measurement interval. Target utilization is a parameter which is set to a fraction (close to, but less than 100%). The load factor, $z$, is an indicator of the congestion level of the link. The optimal operating point is at an overload value equal to one.

The fair share of each VC, $FairShare$, is also computed as follows:

$$\text{FairShare} \leftarrow \frac{\text{ABR Capacity}}{\text{Number of Active Connections}}$$

The switch allows each connection sending at a rate below the $FairShare$ to rise to $FairShare$. If the connection does not use all of its $FairShare$, then the switch fairly allocates the remaining capacity to the connections which can use it. For this purpose, the switch calculates the quantity:

$$\text{VCShare} \leftarrow \frac{CCR}{z}$$

If all VCs changed their rate to their $VCShare$ values then, in the next cycle, the switch would experience unit overload ($z = 1$). $VCShare$ aims at bringing the system to an efficient operating point, which may not necessarily be fair. A combination of the $VCShare$ and $FairShare$ quantities is used to rapidly reach optimal operation as follows:

$$\text{ER Calculated} \leftarrow \text{Max (FairShare, VCShare)}$$

The calculated ER value cannot be greater than the ABR Capacity which has been measured earlier. Hence, we have:

$$\text{ER Calculated} \leftarrow \text{Min (ER Calculated, ABR Capacity)}$$

To ensure that the bottleneck ER reaches the source, each switch computes the minimum of the ER it has calculated as above and the ER value in the RM cell, and indicates this value in the ER field of the RM cell.

The algorithm described above is the main algorithm, but several other steps are carried out to avoid transient overloads and variations in measurement, and drain the transient queues. Moreover, the algorithm is modified to achieve max-min fairness as described in sections 5 and 6.

# 4 The Measurement Interval

ERICA measures the required quantities over consecutive intervals and uses the measured quantities in each interval to calculate the feedback in the next interval. The length of the measurement interval limits the amount of variation which can be eliminated. It also determines how quickly the feedback can be given to the sources, because ERICA gives the same feedback value per source during each measurement interval. Longer intervals produce better averages, but slow down the rate of feedback.

The ERICA algorithm estimates the number of active VCs to use in the computation of the fair share by considering a connection active if the source sends *at least one cell during the measurement interval.* This can be inaccurate if the source is sending at a low rate and the measurement interval is short. In this paper, we propose a better method for estimating the number of active connections. The new method is not as sensitive to the length of the measurement interval. It also eliminates the need to perform some of the steps of the ERICA algorithm, as described in the next section.

# 5 ERICA Fairness Solution

Assuming that the measurements do not exhibit high variation, the original ERICA algorithm converges to efficient operation in all cases. The convergence from transient conditions to the desired operating point is rapid, often taking less than a round trip time. We have, however, discovered cases in which the original algorithm does not converge to max-min fair allocations. This happens if all of the following three conditions are met: (1) the load factor $z$ becomes one, (2) there are some connections which are bottlenecked upstream, (3) the source rate for all remaining connections is greater than the $FairShare$. In this case, the system remains in its current state, because the term $CCR/z$ is greater than $FairShare$ for the non-bottlenecked connections.

This problem was overcome in ERICA as follows. The algorithm is extended to remember the highest allocation made during each measurement interval, and ensure that all eligible connections can also get this high allocation. To do this, $MaxAllocPrevious$ stores the maximum allocation given in the previous interval. For $z > 1 + \delta$, where $\delta$ is a small fraction, we use the basic ERICA algorithm and allocate Max (FairShare, VCShare). But, for $z \leq 1+\delta$, we attempt to make all the rate allocations equal, by assigning ER to Max (FairShare, VCShare, MaxAllocPrevious). The aim of introducing the quantity $\delta$ is to force the allocation of equal rates when the overload is fluctuating around unity, thus avoiding unnecessary rate oscillations. The remainder of this paper proposes a more accurate method to compute the max-min fair shares for all the contending connections.

# 6 An Accurate Method to Determine the Fair Bandwidth Share

As previously discussed, ERICA determines the number of active connections by considering a source as active if at least one cell from this source is sent during the measurement interval. A more accurate method to compute activity and eliminate the need for the proposed solution to the fairness problem is to compute a quantity that we call the "effective number of active VCs" and use this quantity to compute the $FairShare$, as described next.

## 6.1 Basic Idea

We redefine the $FairShare$ quantity to be ***the maximum share a VC could get at this switch under max-min fairness criteria***. Hence, the $FairShare$ is calculated as follows:

$$\text{FairShare} = \frac{\text{ABR capacity}}{\text{Effective number of active VCs}}$$

The main innovation is the computation of the effective number of active VCs. The value of the effective number of active VCs depends on the activity level of each of the VCs. The activity level of a VC is defined as follows:

$$\text{Activity level} = \text{Min}(1, \frac{\text{Source Rate}}{FairShare})$$

Thus, VCs that are operating at or above the $FairShare$ are each counted as one. The VCs that are operating below the $FairShare$ (and are probably not bottlenecked at this switch) only contribute a fraction. The VCs that are bottlenecked at this switch are considered fully active while other VCs are considered partially active.

The effective number of active VCs is the sum of the activity levels for all VCs:

$$\text{Effective number of active VCs} = \sum_i \text{Activity level of VC}_i$$

Note that the definition of activity level depends upon the $FairShare$, and the definition of the $FairShare$ depends upon the activity levels. Thus, the definitions are recursive.

## 6.2 Examples of Operation

**Example 1 (stability):**

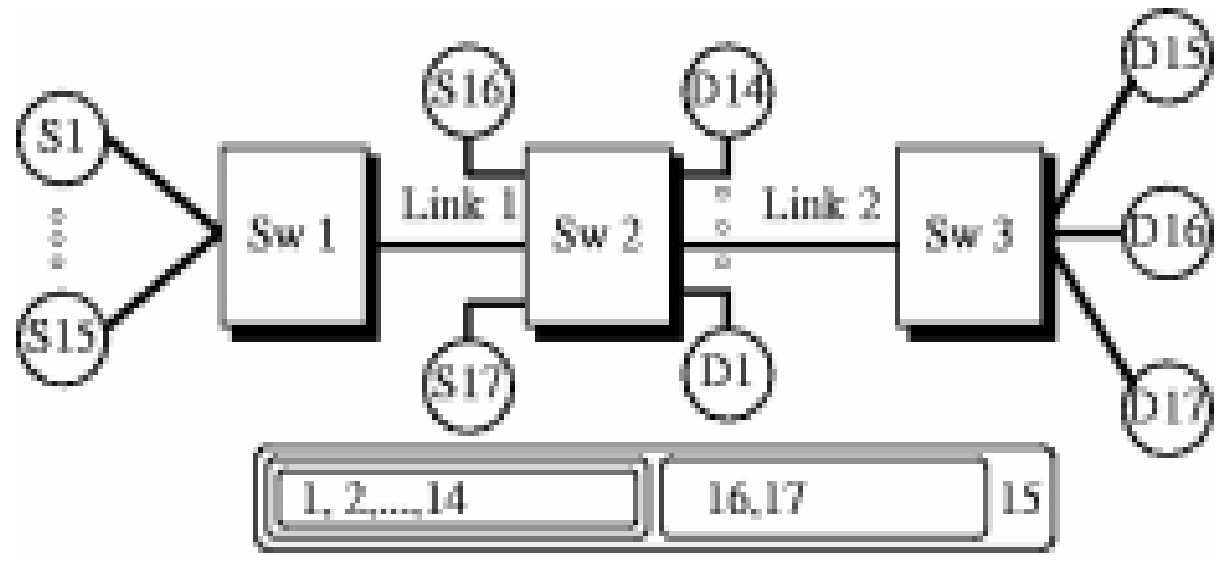

Figure 2: Upstream Configuration

Consider the upstream bottleneck case with 17 VCs shown in figure 2. It has been shown in [6] that this configuration demonstrates the unfairness of the original ERICA algorithm as described in section 3, which necessitates the addition described in section 5.

Assume that the target capacity is 150 Mbps. For the second switch, when the rates for ($S1$, $S16$, $S17$) are (10, 70, 70):
Iteration 1: Assume FairShare = 70 Mbps
Activity = (10/70, 70/70, 70/70) = (1/7, 1, 1)
Effective number of active VCs = 1 + 1 + 1/7 = 15/7
Iteration 2: FairShare = Target capacity/Effective number of active VCs = 150/2.14 = approximately 70 Mbps

Hence, this example shows that the system is stable at the allocation of (10, 70, 70). At any other allocation, the scheme will calculate the appropriate $FairShare$ that makes the allocation eventually reach this point, as seen in the next two examples.

**Example 2 (rising from a low FairShare):**

For the same configuration, when the rates are (10, 50, 90):
Iteration 1: Assume FairShare = 50 Mbps
Activity = (10/50, 50/50, 1) = (0.2, 1, 1)
Effective number of active VCs = 0.2 + 1 + 1 = 2.2
Iteration 2: FairShare = 150/2.2 = approximately 70 Mbps

Again, the scheme reaches the optimal allocation within a few round trip times.

**Example 3 (dropping from a high FairShare):**

For the same configuration, when the rates are (10, 50, 90):
Iteration 1: Assume FairShare = 75 Mbps
Activity = (10/75, 50/75, 1) = (0.13, 0.67, 1)
Effective number of active VCs = 0.13 + 0.67 + 1 = 1.8
Iteration 2: FairShare = 150/1.8 = 83.3 Mbps

Suppose the sources start sending at the new rates, except for the first one which is bottlenecked at 10 Mbps. Also assume that FairShare is still at 83.3 Mbps.

Activity = (10/83.3, 83.3/83.3, 83.3/83.3) = (0.12, 1, 1)
Effective number of active VCs = 0.12 + 1 + 1 = 2.12
FairShare = 150/2.12 = approximately 70 Mbps

Again, the scheme reaches the optimal allocation after the sources start sending at the specified allocations, which is within a few round trip times.

### 6.3 Derivation

The following derivation shows how we have verified the correctness of our method of calculation of the number of active connections. The new algorithm is based upon some of the ideas presented in the MIT scheme [3, 2]. The derivation depends on classifying active VCs as either underloading VCs or overloading VCs. A VC is *overloading* if it is bottlenecked at this switch; otherwise the VC is said to be *underloading*. In the MIT scheme, a VC is determined to be overloading by comparing the computed $FairShare$ value to the desired rate indicated by the VC source. In our scheme, we classify a VC as overloading if its source rate is greater than the $FairShare$ value. Our algorithm only performs one iteration every measurement interval, and is not of the complexity of the order of the number of VCs, as with the MIT scheme.

The MIT scheme has been proved to compute max-min fair allocations for connections within a certain number of round trips (see the proof in [2]). We prove that the MIT scheme reduces to our equation as follows. According to the MIT scheme:

$$FairShare = \frac{\text{ABR Capacity} - \sum_{i=1}^{N_u} Ru_i}{N - N_u}$$

where:
$Ru_i$ = Rate of $i^{th}$ underloading source ($1 \leq i \leq N_u$)
$N$ = Total number of VCs
$N_u$ = Number of underloading VCs

Substituting $N_o$ for the denominator term, this becomes:

$$\text{FairShare} = \frac{\text{ABR Capacity} - \sum_{i=1}^{N_u} Ru_i}{N_o}$$

where:
$N_o$ = Number of overloading VCs ($N_u + N_o = N$)

Or:

$$FairShare \times N_o + \sum_{i=1}^{N_u} Ru_i = \text{ABR Capacity}$$

Factoring $FairShare$ out in the left hand side:

$$FairShare \times (N_o + \sum_{i=1}^{N_u} \frac{Ru_i}{FairShare}) = \text{ABR Capacity}$$

Or:

$$FairShare = \frac{\text{ABR Capacity}}{N_o + \sum_{i=1}^{N_u} \frac{Ru_i}{FairShare}}$$

Substituting $N_{eff}$, we get:

$$FairShare = \frac{\text{ABR Capacity}}{N_{eff}}$$

where:

$$N_{eff} = N_o + \sum_{i=1}^{N_u} \frac{Ru_i}{FairShare}$$

This means that the effective number of active VCs is equal to the number of overloading sources, plus the fractional activity of underloading sources.

## 7 Performance Analysis

The new algorithm has been tested for a variety of networking configurations using several performance metrics. The results were similar to the results obtained with the ERICA algorithm [6], except that the new algorithm is max-min fair (without executing the steps described in section 5), and is less sensitive to the length of the measurement interval. A sample of the results demonstrating fairness is described in this section.

### 7.1 Parameter Settings

Throughout our experiments, the following parameter values are used:

1. All links have a bandwidth of 155.52 Mbps.
2. All links are 1000 km long.
3. All VCs are bidirectional.
4. The source parameter Rate Increase Factor (RIF) is set to one, to allow immediate use of the full explicit rate indicated in the returning RM cells at the source.
5. The source parameter Transient Buffer Exposure (TBE) is set to large values to prevent rate decreases due to the triggering of the source open-loop congestion control mechanism. This was done to isolate the rate reductions due to the switch congestion control from the rate reductions due to TBE.
6. The switch target utilization parameter was set to 90%. This factor is used to scale down the ABR capacity term used in the ERICA algorithm. Alternatively, a queue control function can be used to achieve a target queuing delay and queue lengths [8].

7. The switch measurement interval was set to the minimum of the time to receive 100 cells and 1 ms.

8. All sources are deterministic, i.e., their start/stop times and their transmission rates are known.

## 7.2 Simulation Results

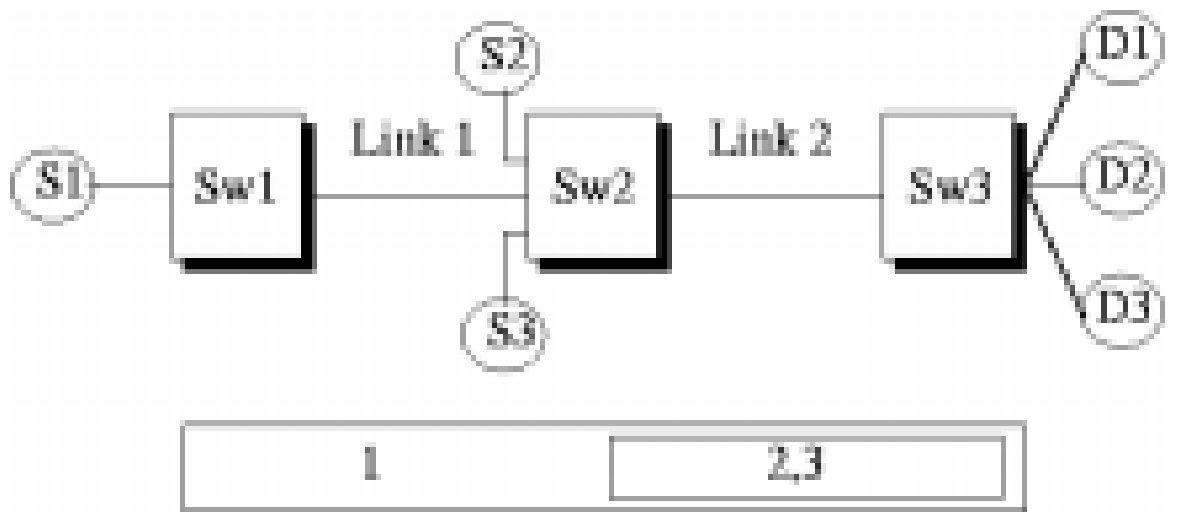

Figure 3: Three source configuration

In order to test fairness, we have simulated a three source configuration where one of the sources is bottlenecked at a low rate (10 Mbps). Hence, even though the network gives that source feedback to increase its rate, it never sends at a rate faster than 10 Mbps. The other two sources start transmission at different ICR values. The aim of the configuration is to examine whether the two non-bottlenecked sources will reach the same ACR values, utilizing the bandwidth left over by the first source. (A number of other configurations was simulated, and the results indicated good performance of the proposed algorithm.)

Figure 3 illustrates the configuration simulated. Note that the round trip time for the $S2$ and $S3$ connections is 30 ms, while that for the $S1$ connection is 40 ms. This configuration is almost identical to the one used in the examples in section 6 (figure 2), except that connection $S1$ to $D1$ is bottlenecked at the source $S1$ itself, and not at "Link 1." The reason we chose to demonstrate a source bottleneck situation here (and not a link bottleneck situation like figure 2) is to demonstrate the effect of using the CCR field in the RM cells versus measuring the source rate.

The results are presented in the form of two graphs for each configuration: (a) Graph of allowed cell rate (ACR) in Mbps over time for each source. (b) Graph of the effective number of active VCs $N_{eff}$ at the bottleneck port.

Figure 4 illustrates the performance of the original ERICA algorithm without the fairness step discussed in section 5. Source $S1$ is the bottlenecked source. Sources $S2$ and $S3$ start sending at different ICR (and hence ACR) values. Their ICR values and that of $S1$ add up to little more than the the link rate, so there is little overload. Observe that the rates of $S2$ and $S3$ remain different, leading to unfairness. The number of active VCs is counted using the original ERICA method, so the switch sees 3 sources (see figure 4(b)), and the $FairShare$ value remains at around 50 Mbps. Hence, the source $S2$ never increases its rate to make use of the bandwidth left over by $S1$ and only $S3$ utilizes this bandwidth.

Figure 5 illustrates how the fairness problem was overcome in ERICA by the change described in section 5. In this case, the sources are given the maximum allocation in case of underload or unit load, and hence all sources get an equal allocation. The modified algorithm is max-min fair.

Figure 6 illustrates the results with the new method to calculate the fair share of the bandwidth. Observe that the allocations are max-min fair in this case, without needing to apply the maximum allocation algorithm as in the previous case. This is because the method of calculation of the effective number of active connections is different. Figure 6 shows that after the initialization period, the effective number of active VCs stabilizes at 1 (for $S2$), plus 1 (for $S3$), plus 10/50 (for $S1$), which gives $1 + 1 + 0.2 = 2.2$ sources. The method also stabilizes to the correct number even *if the length of the measurement interval is short*, unlike the original method where the length of the measurement interval must be long enough to detect cells from all sources, even low-rate sources.

The proposed method works correctly for all cases when there are *link bottlenecks* at various locations (e.g., the configuration in figure 2), since it correctly calculates the activity level of each connection based on its CCR value. However, observe that in *source bottleneck* cases, the CCR value cannot be simply obtained from the forward RM cells, but must be measured by the switches. This is because, in source bottleneck situations, the source indicates its ACR value in the CCR field of the RM cell, but the source may actually be sending at a much lower rate than its ACR.

For example, for the configuration discussed above (figure 3), assume that we were relying on the CCR values in the RM cells. Figure 7 shows that the new method is not fair in this case, since source $S1$ indicates an ACR of 50 Mbps so the effective number of active connections stabilizes at 3 (see figure 7(b)), and the $FairShare$ remains at 50 Mbps. But source $S1$ is only sending at 10 Mbps. CCR measurement at the switch detects this, and hence arrives at the correct allocation as seen in figure 6.

## 7.3 Observations on the Results

From the simulation results, we can make the following observations about the performance of the proposed algorithm:

- During transient phases, if the $FairShare$ value increases, the $N_{eff}$ value decreases (since it uses the $FairShare$ value in the denominator), and $FairShare$ further increases (since it uses $N_{eff}$ in the denominator), so $N_{eff}$ further decreases, and so on, until the correct values of rates, $N_{eff}$ and $FairShare$ are reached. Then the proposed scheme is provably fair and efficient in steady state (see figure 6(a) and (b)). Although the scheme is recursive, its transient response was found to be very fast.

- Even if the measurement interval is so short such that no cells are seen from many low-rate sources, the proposed

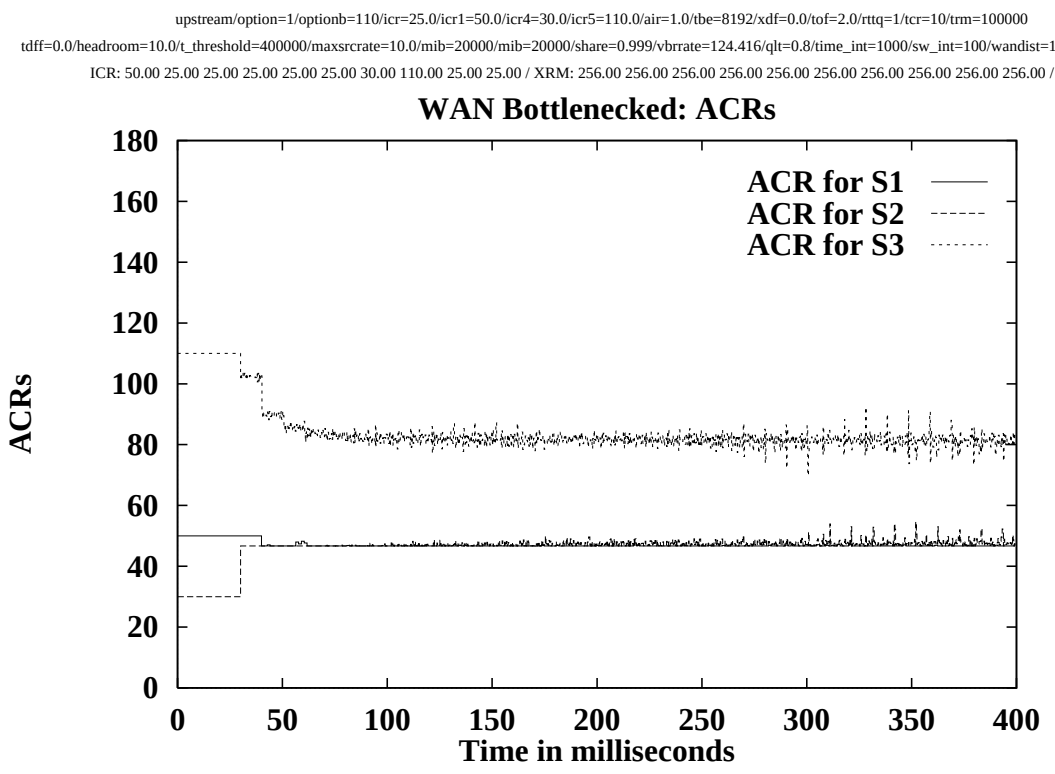

(a) Allowed Cell Rate

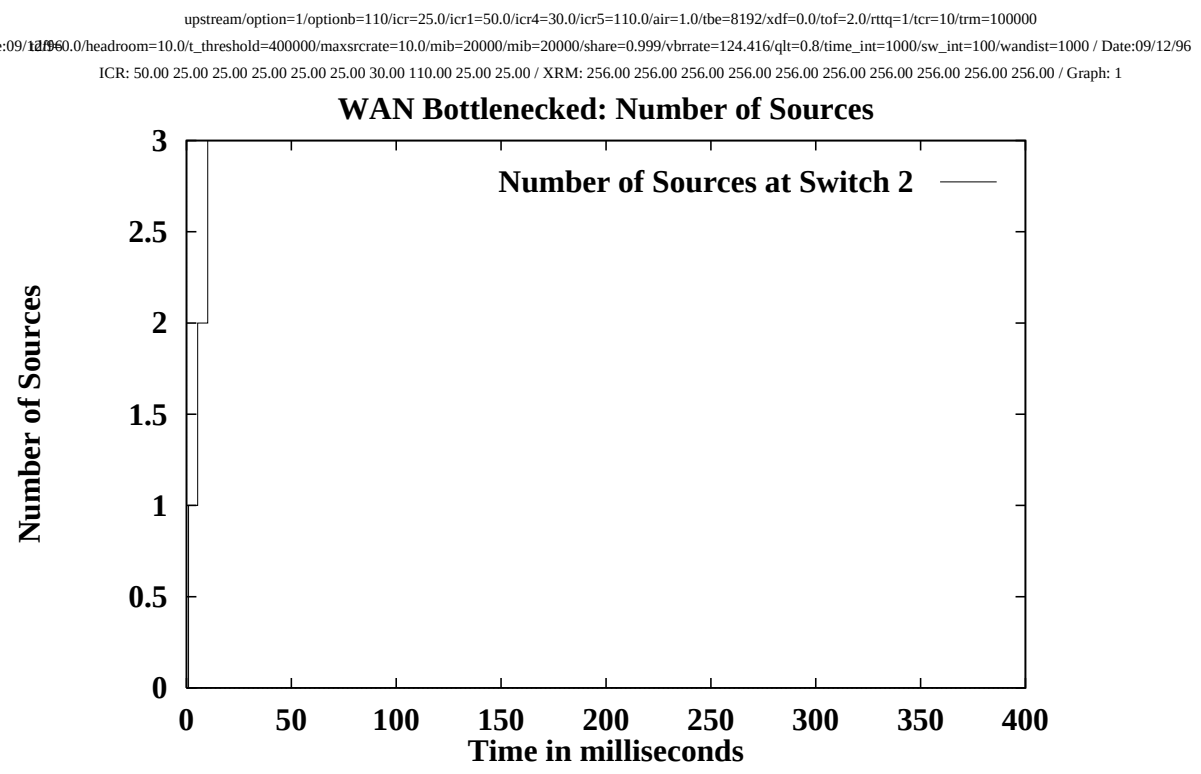

(b) Number of Active VCs

Figure 4: Results for a WAN three source bottleneck configuration with the original ERICA

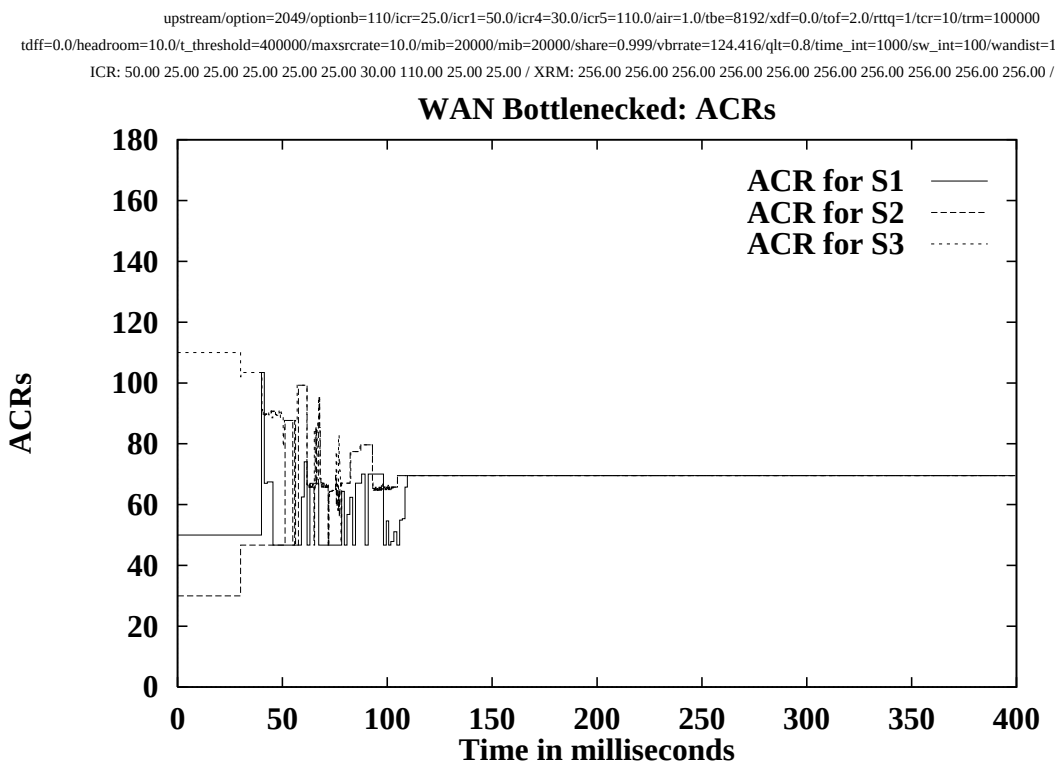

(a) Allowed Cell Rate

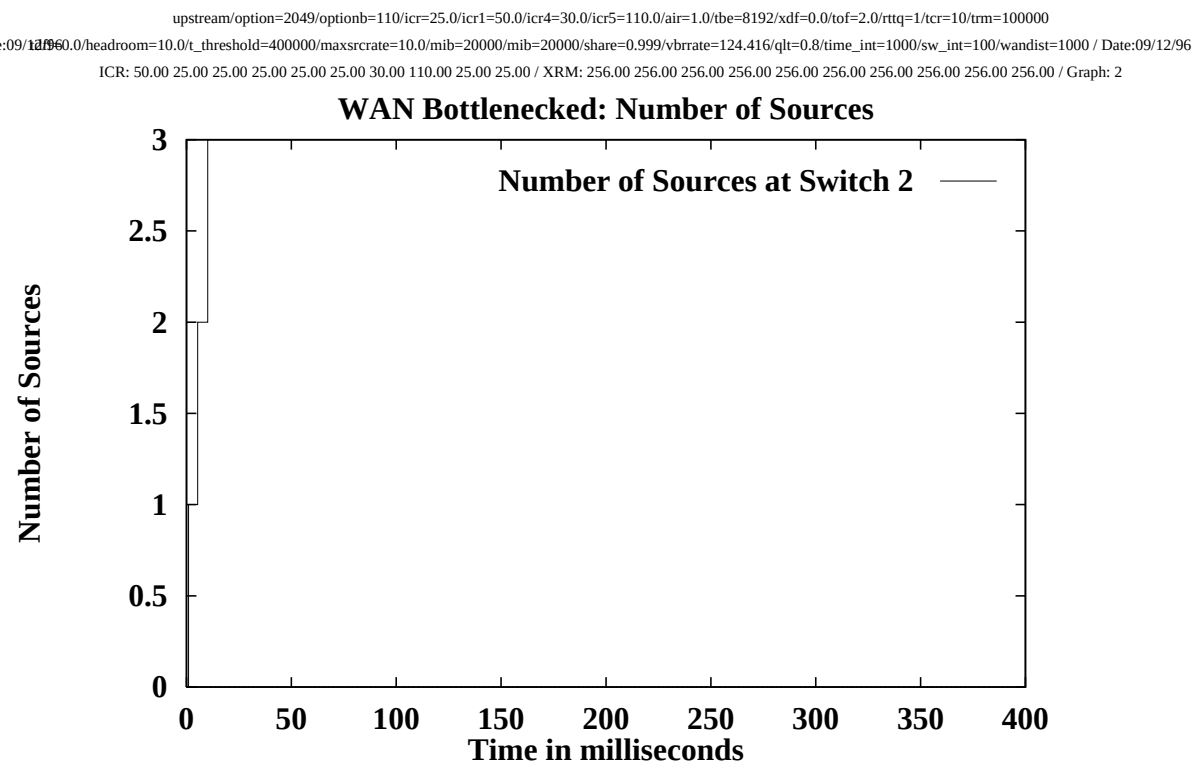

(b) Number of Active VCs

Figure 5: Results for a WAN three source bottleneck configuration with ERICA

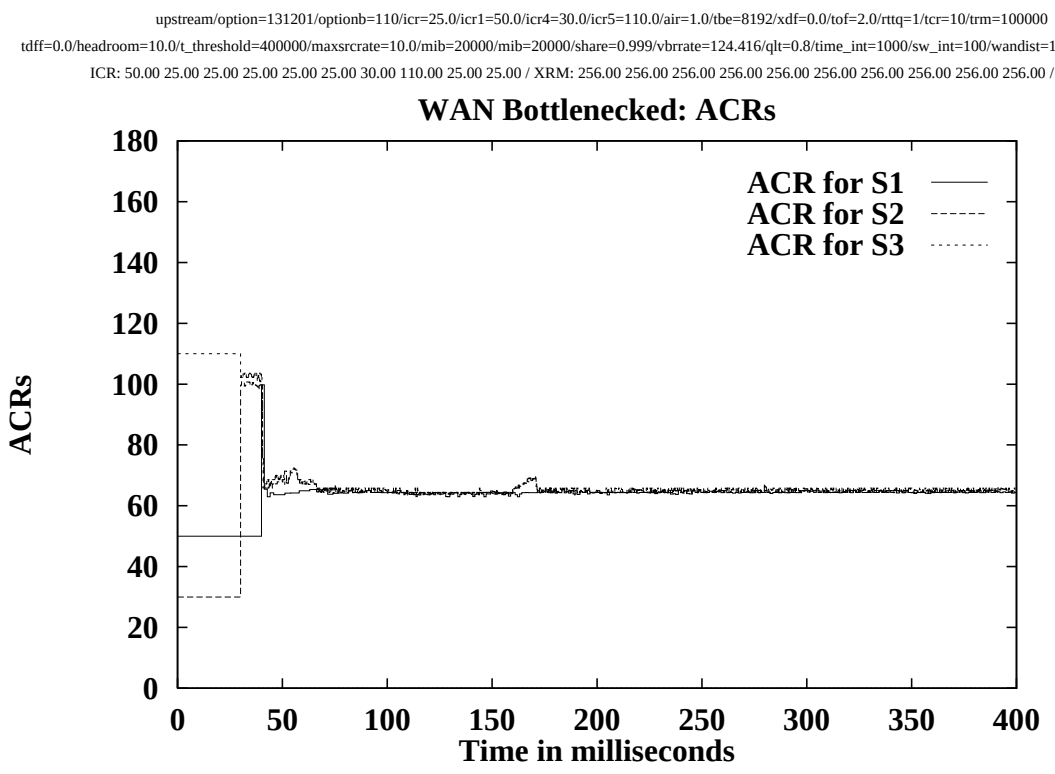

(a) Allowed Cell Rate

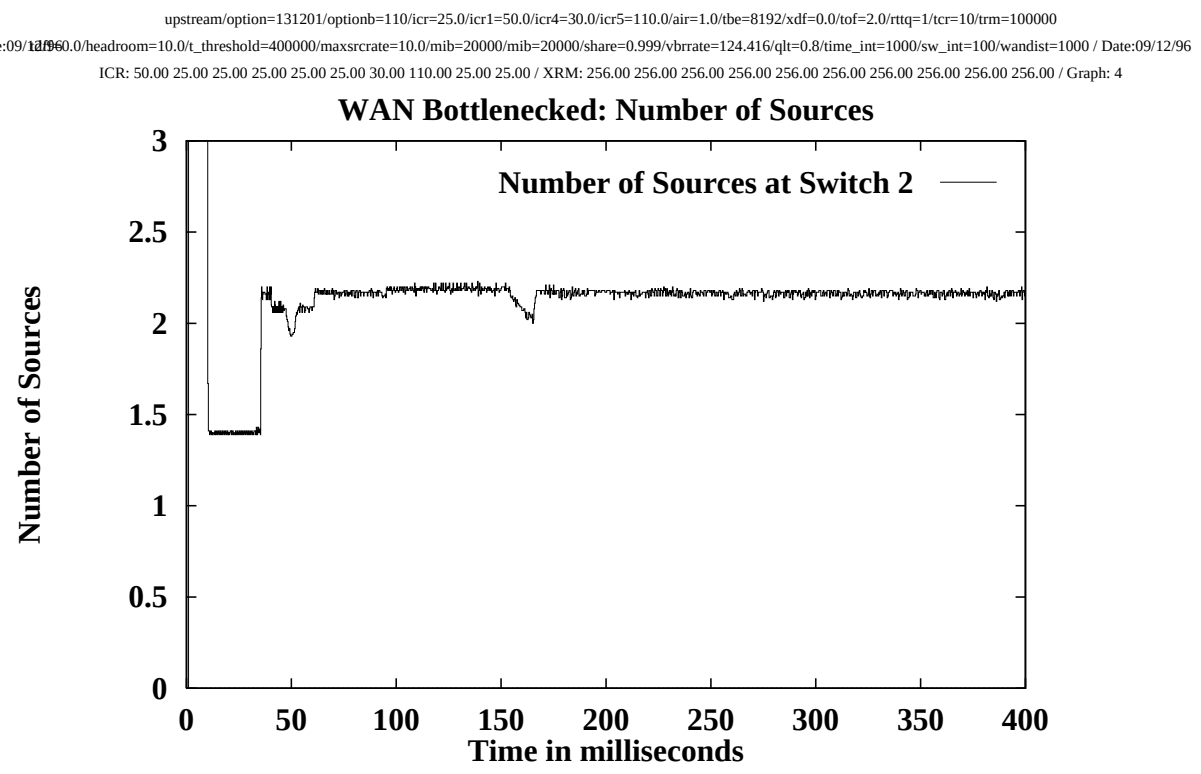

(b) Number of Active VCs

Figure 6: Results for a WAN three source bottleneck configuration with the proposed ERICA and source rate measurement at the switch

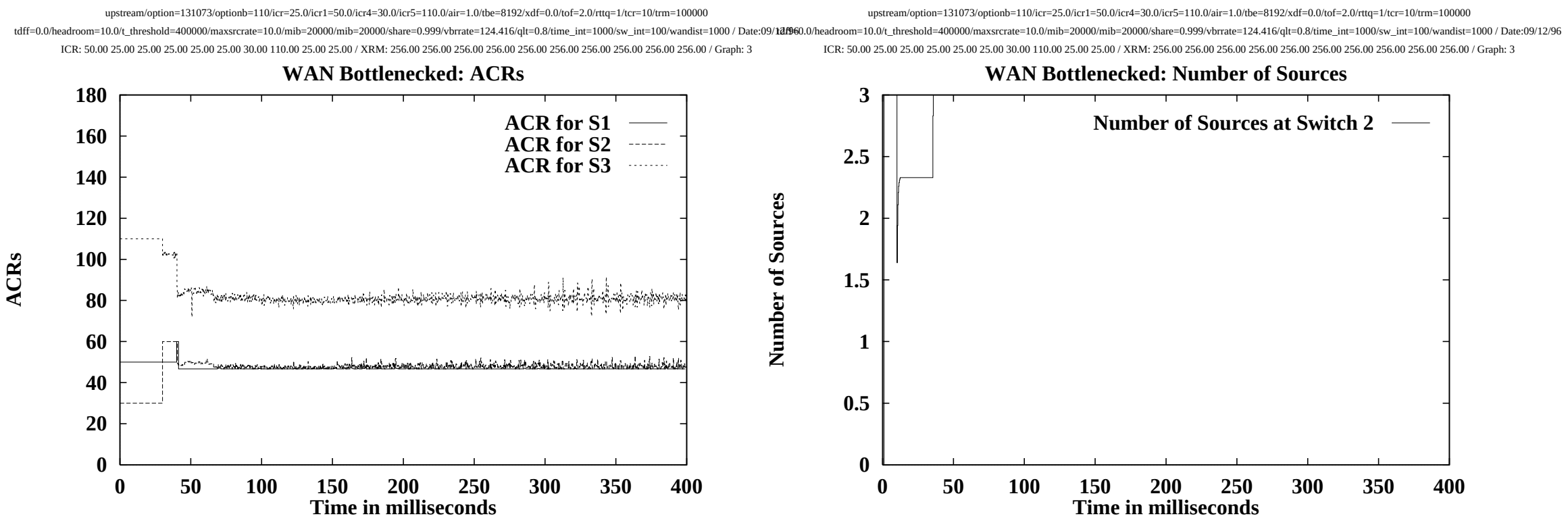

(a) Allowed Cell Rate

(b) Number of Active VCs

Figure 7: Results for a WAN three source bottleneck configuration with the proposed ERICA

method can compute the $FairShare$ of the bandwidth correctly (this result is not shown by the simulations in this paper).

- Without source rate measurement at the switch for each VC, the value of $N_{eff}$ depends on the source ACR, which is not the same as the source rate for source bottleneck cases. Thus, $N_{eff}$ is too large in those cases, and the $FairShare$ term is less than the CCR by Overload term, leading to unfairness. With per-VC source rate measurement, the value of $N_{eff}$ is correct.

# 8 Summary

This paper has proposed and demonstrated a new method to compute the fair bandwidth share for ABR connections in ATM networks. The method relies on distinguishing among underloading connections and overloading connections, and computing the value of the "effective number of active connections." The available bandwidth is divided by the effective number of active connections to obtain the fair bandwidth share of each connection.

The method is provably max-min fair, and can be used to ensure the efficiency and fairness of bandwidth allocations. Integrating this method into ERICA tackles the fairness and measurement interval problems of ERICA, while maintaining the fast transient response, queuing delay control, and simplicity of the ERICA scheme.

Analysis and simulation results were used to investigate the performance of the method. From the results, it is clear how the method overcomes the fairness problem with the original ERICA, as well as its excessive sensitivity to the length of the measurement interval.